\documentclass[twocolumn]{article}
\usepackage{float}
\usepackage{graphicx}
\usepackage{xcolor}

\usepackage[top=2cm, bottom=1.5cm, left=1.5cm, right=1.5cm]{geometry}
\usepackage{titling}
\usepackage{dirtytalk} 
\usepackage{amsmath}

\usepackage[super,numbers,sort&compress]{natbib}
\usepackage{url}
\usepackage{breakurl}
\usepackage[breaklinks]{hyperref}

\usepackage{enumitem}

\usepackage{authblk}

\usepackage{midfloat}
\usepackage{wrapfig}

\usepackage{caption}
\usepackage{setspace}

\usepackage{titlesec}

\usepackage{xfrac}

\titleformat*{\section}{\small\bfseries}
\titleformat*{\subsection}{\small\bfseries}

\title{Cooperation in a fluid swarm of fuel-free micro-swimmers}

\author[1,2]{Matan Yah Ben Zion\thanks{matanbz@gmail.com}}
\author[1]{Yaelin Caba}
\author[1]{Alvin Modin}
\author[1]{Paul M Chaikin}
\affil[1]{\small{Center for Soft Matter Research, Department of Physics, New York University, 726 Broadway Avenue, New York, NY 10003, USA} }
\affil[2]{\small{UMR Gulliver 7083 CNRS, ESPCI Paris, PSL Research University, 10 rue Vauquelin, Paris 75005, France}}

\begin{document}

\maketitle

{\bf Cooperation is vital for the survival of a swarm \cite{Hamann2018}. Large scale cooperation allows murmuring starlings to outmaneuver preying falcons \cite{Procaccini2011}, shoaling sardines to outsmart sea lions \cite{BBC2015}, and homosapiens to outlive their Pleistocene peers \cite{Harari2015}. On the micron-scale, bacterial colonies show excellent resilience thanks to the individuals' ability to cooperate even when densely packed, mitigating their internal flow pattern to mix nutrients, fence the immune system, and  resist antibiotics \cite{Costerton1999,Saintillan2008,Secchi2016,Marchetti2013,Dunkel2013,Wensink2012,Zhang2010c,Dombrowski2004,Xu2019,Gompper2020}. Production of an artificial swarm on the micro-scale faces a serious challenge --- while an individual bacterium has an evolutionary-forged internal machinery to produce propulsion, until now, artificial micro-swimmers relied on the precise chemical composition of their environment to directly fuel their drive \cite{Gompper2020, VanDerLinden2019, Geyer2019, Bauerle2018, Yan2016, Lozano2016, Maass2016, Izri2014, Howse2007, Palacci2013}. When crowded, artificial micro-swimmers compete locally for a finite fuel supply, quenching each other's activity at their greatest propensity for cooperation. Here we introduce an artificial micro-swimmer that consumes no chemical fuel and is driven solely by light. We couple a light absorbing particle to a fluid droplet, forming a colloidal chimera that transforms light energy into propulsive thermo-capillary action. The swimmers' internal drive allows them to operate and remain active for a long duration (days) and their effective repulsive interaction allows for a high density fluid phase. We find that above a critical concentration, swimmers form a long lived crowded state that displays internal dynamics. When passive particles are introduced, the dense swimmer phase can re-arrange and spontaneously corral the passive particles. We derive a geometrical, depletion-like condition for corralling by identifying the role the passive particles play in controlling the effective concentration of the micro-swimmers.}

\begin{singlespace}
\begin{figure}[h!]
\centering
\includegraphics[width=0.5\textwidth]{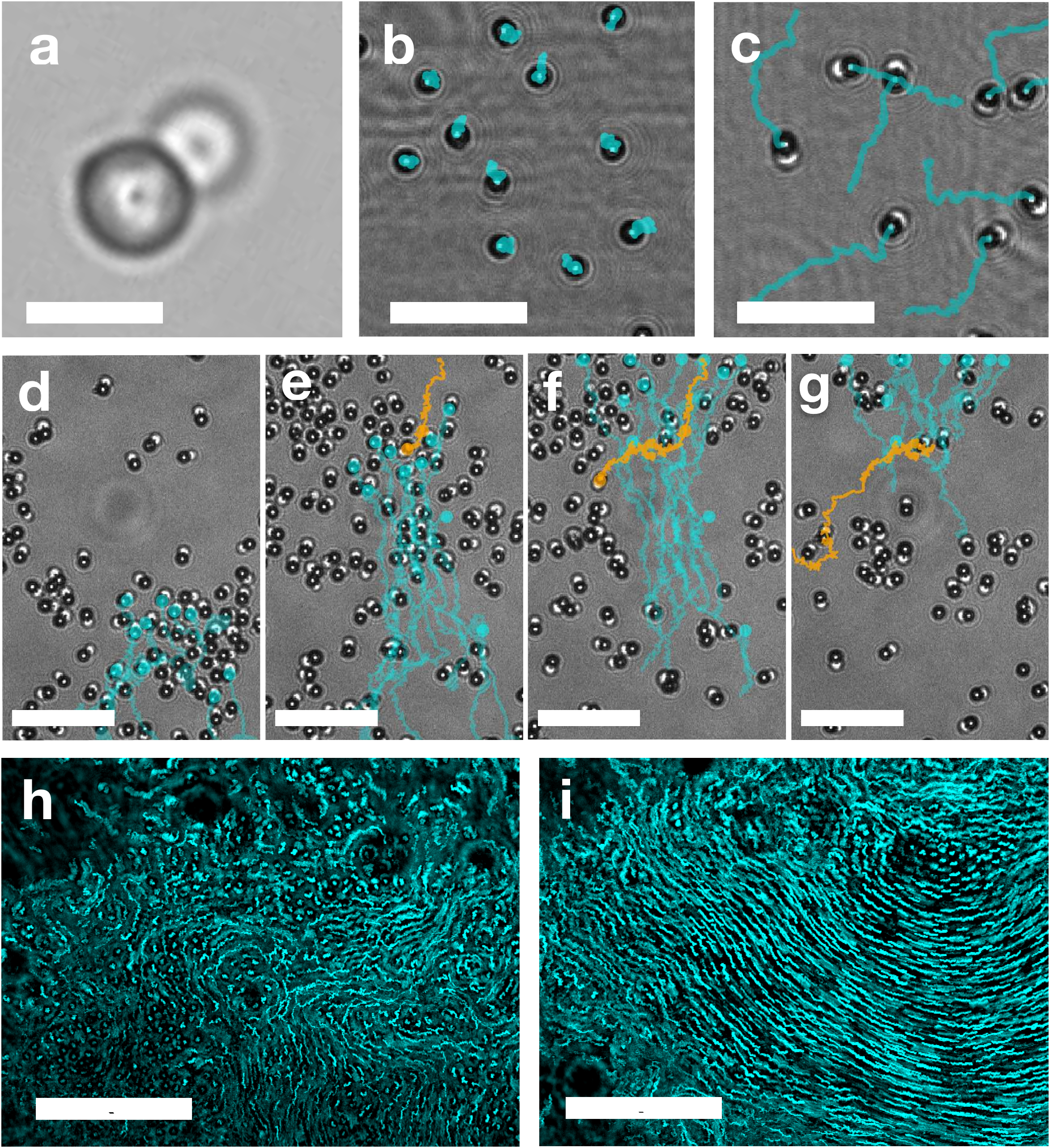} 
\caption{The light driven fuel-free micro-swimmers display distinct behaviours at different densities. {\bf a} a single micro-swimmer is composed of a light absorbing particle (dark sphere), and a fluid droplet (white sphere). Individual swimmers transition from diffusive dynamics when light is off ({\bf b}) to more ballistic when light is turned on ({\bf c}) (see Supplementary Video 3). {\bf d} at an intermediate concentration, swimmers collide, align, and create small co-moving groups. {\bf d}-{\bf g} an individual swimmer (orange) maintains its activity even when passing through an oppositely moving group (cyan), penetrating, and emerging from the other side (see Supplementary Video 4). {\bf h}-{\bf i} long exposure (1 min) images of swimmers at high concentrations reveal internal flow structures (see Supplementary Video 5). Scale bars: {\bf a}: $5\;\mu\rm{m}$; {\bf b}, {\bf c}: $20\;\mu\rm{m}$; {\bf d}-{\bf g}: $25\;\mu\rm{m}$; {\bf} h - {\bf i}: $50\;\mu\rm{m}$.}
\vspace{-0.75cm}
\label{figSwimmersPhases}
\end{figure}

 \end{singlespace}

\begin{singlespace}
\begin{figure}[!hb]
\centering
\includegraphics[width=0.5\textwidth]{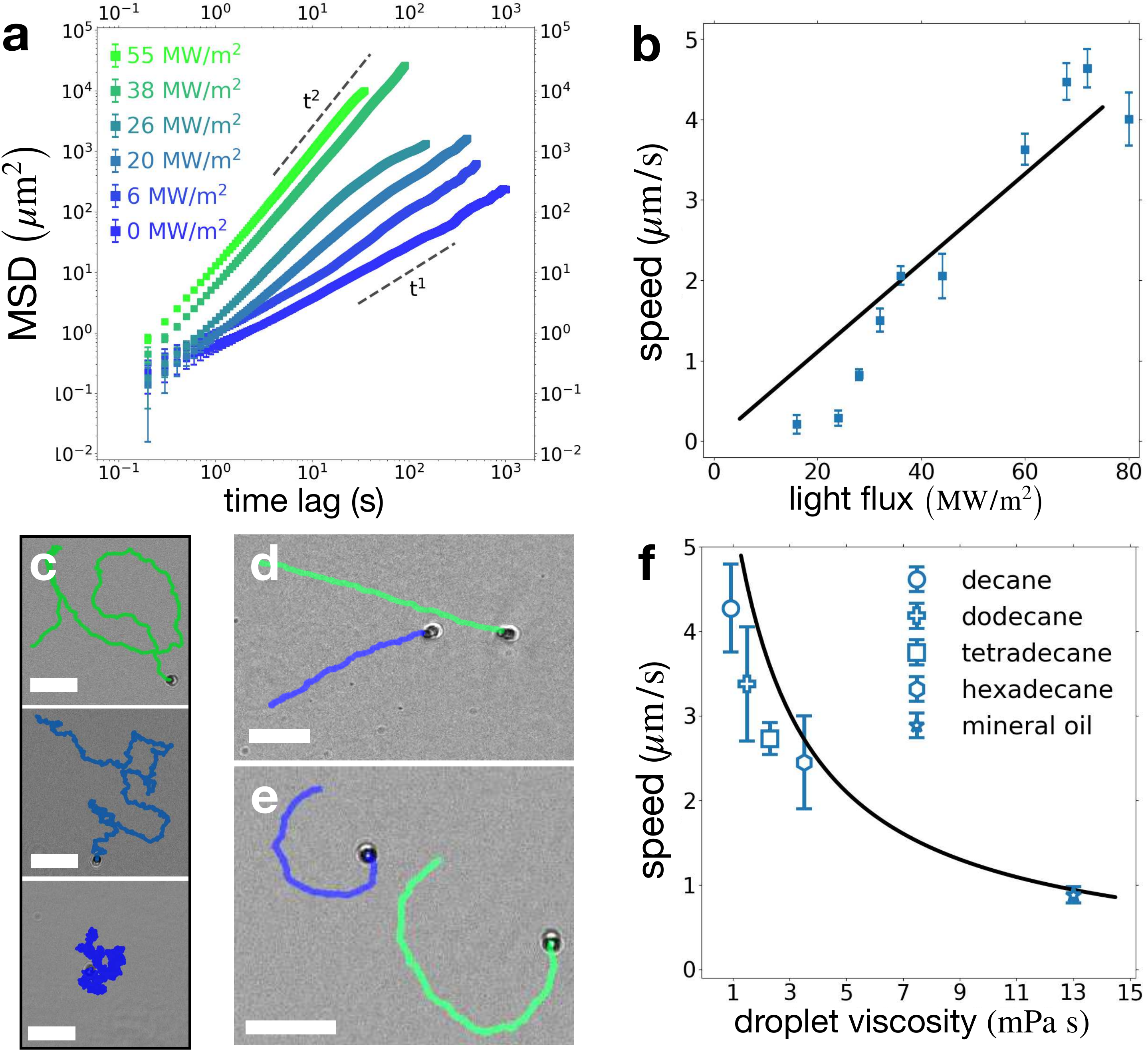} 
\caption{The internal drive of the fluid micro-swimmers relies on light activated thermo-capillary action, independent from a chemical fuel in the surrounding bath. {\bf a} the mean square displacement (MSD) shows an individual swimmer’s motility can be continuously tuned from diffusive ($\propto t^1$) to ballistic ($\propto t^2$). {\bf b} the speed of a swimmer scales linearly with the power flux of the light source reaching more than one body length per second. {\bf c} with increasing power flux, trajectories of free swimmers show transition from random (bottom) to more ballistic (top) motion. Using an external magnetic field the orientation of a swimmer can be controlled to move on a straight line ({\bf d}) or in circles ({\bf e}) allowing a direct measurement of the nominal swimming speed. {\bf f} increasing the internal viscosity of the fluid droplet directly reduces the speed of the swimmer. Solid line is the theoretically predicted swimming speed using no fitting parameters (full characterisation of optical absorption and thermo-capillary measurements are found in Supplementary Information section \ref{supSwimmerModel}). Scale bars: $20\;\mu\rm{m}$.}
\label{figSingleSwimmer}
\end{figure}
\end{singlespace}

In order to explore the dynamics of a dense swarm of micro-swimmers, we synthesized a colloidal dimer by coupling a light absorbing particle (M280) to a liquid droplet (n-dodecane), forming a $5\;\mu\rm{m}$ long, peanut shaped swimmer (see Fig.\ref{figSwimmersPhases}a and in Methods for synthesis steps\cite{BenZion2020}). When exposed to light, dimers commence a persistent swimming motion; when light is turned off, the dimers revert to random Brownian motion (see Fig.\ref{figSwimmersPhases}b,c). As a light source, we used a wide beam (diameter $d = 310\; \mu \rm{m}$) infrared (wavelength $\lambda = 1064\;\rm{nm}$) laser (see section \ref{supLaserHeating} in Supplementary Information for details). Typical experiments were performed in deionized water, yet swimmers are also active in saline phosphate buffer (PBS), or when exposed to a visible, non-coherent, light sources (see Methods for experimental setup and sample preparation). We find drastically different collective dynamics of the active particles when their area fraction, $\phi_A$, is increased, transitioning from quickly dispersing when dilute ($\phi_A < 0.14$, Fig.\ref{figSwimmersPhases}b-c and Supplementary Video 3); to forming transient, aligned aggregates at intermediate concentration ($0.14<\phi_A < 0.28$, Fig.\ref{figSwimmersPhases}d-g, and Supplementary Video 4); Finally, at high concentrations, swimmers crowd in a dense, active colony with internal flows ($0.48\le\phi_A$, Fig.\ref{figSwimmersPhases}h-i, and Supplementary Video 5).

\begin{figure*}[h!]
\centering
\includegraphics[width=0.8\textwidth]{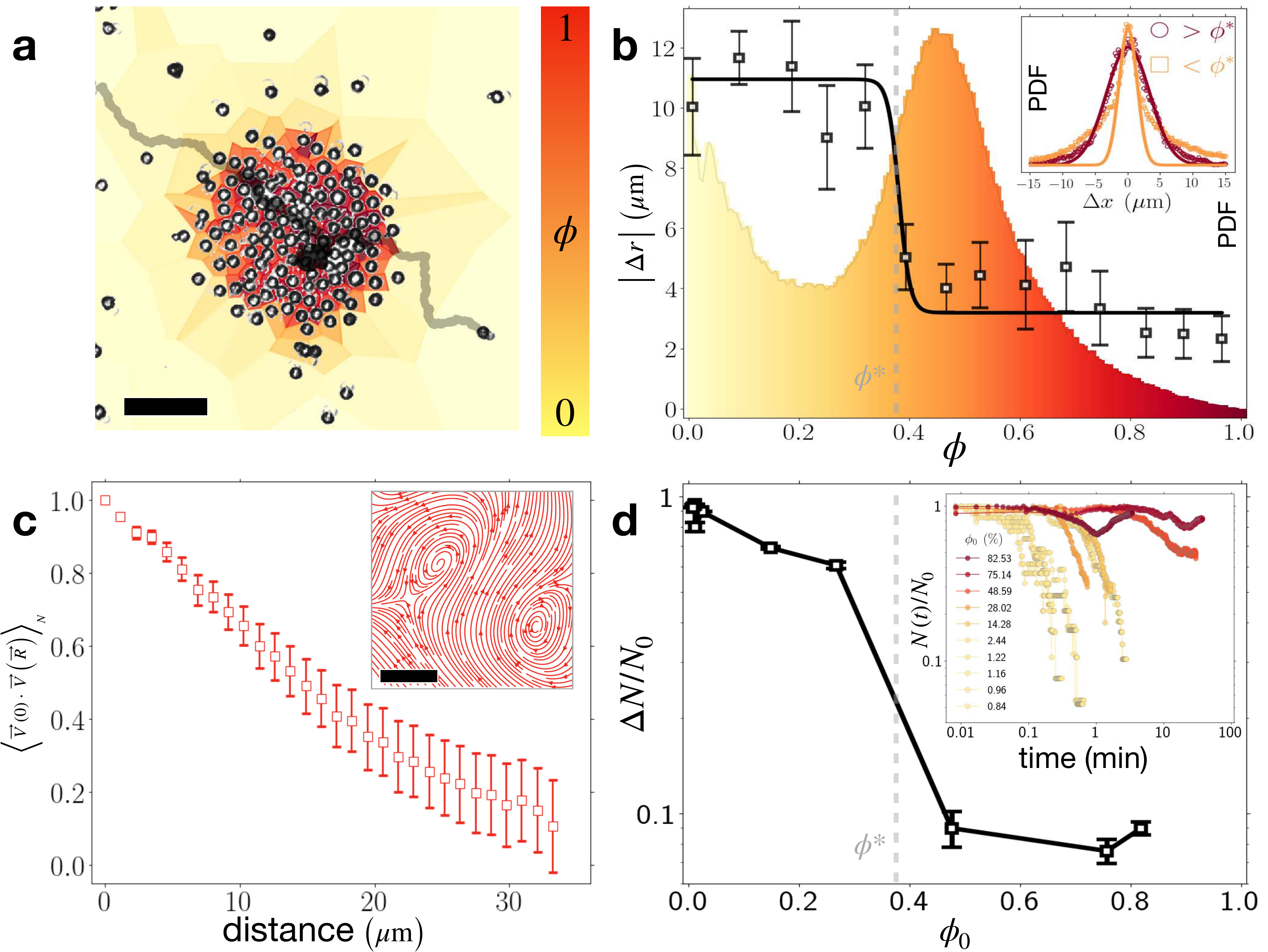} 
\caption{At sufficiently high average area fraction, $\phi_0$, the swimmer population segregates into a long lived state of dilute and dense fluids. {\bf a} the instantaneous local density of each swimmer is measured from the inverse size of the corresponding Voronoi cell. Both the dilute and the dense phases are fluid as can be seen from the trace of an individual swimmer crossing the different phases, moving slower (darker trace) in the denser region (see Supplementary Video 1). {\bf b} the probability for finding an individual swimmer at specific local density shows two characteristic phases. Particle displacement over a fixed interval (5 seconds) shows a sharp decrease above local concentration of $\phi^*=0.38$ (black squares). Inset shows the displacement probability distribution function for an individual swimmer below (yellow squares) and above (red circles) $\phi^*$. The low density region deviates from a Gaussian distribution (yellow curve) while the displacement PDF in the dense phase fits a Gaussian distribution (red curve). {\bf c} velocity correlation function within the dense phase indicates flows correlate on a length scale much larger than the swimmer size. Inset shows streamlines measured using particle image velocimetry, displaying an internal flow structure (see section \ref{supPIV} in Supplementary Information). {\bf d} at lower initial concentration, $\phi_0\le0.28$, most of the swimmers will disperse, $\Delta N \gg N_0$, within a few seconds up to minutes. At higher initial concentrations, $\phi_0\ge0.49$, the swarm forms a persistent crowded state, which remains throughout the measurement time (hours). Inset shows temporal evolution. Scale bars are $20\;\mu\rm{m}$.}
\label{figMIPS}
\end{figure*}

In both the intermediate and high concentration regimes individual swimmers are often observed to swim within, and sometimes against, a group of other active particles (see Fig.\ref{figSwimmersPhases}d-f, Fig.\ref{figMIPS}a, and Supplementary Videos 1 and 4). This high degree of autonomy has been observed in living bacteria \cite{Dombrowski2004,Zhang2010c,Wensink2012,Dunkel2013,Xu2019,Gompper2020}, but was not reported in previously studied systems of artificial micro-swimmers fuelled by either external ionic currents \cite{VanDerLinden2019,Geyer2019}, consumption of micelles \cite{Maass2016}, hydrolysis of hydrogen peroxide \cite{Palacci2013}, or critical de-mixing of their surrounding buffer \cite{Bauerle2018}. Instead, those systems exhibit interrupted internal dynamics, with the crowded phase being frozen, and sometimes crystalline.

The propulsion mechanism of a single active particle is quantitatively captured using a thermo-capillary drive for a fluid droplet at a local temperature gradient\cite{Young1959,Barton1989}. Since surface tension, $\gamma$, is a powerful agent on the micro-scale and is generally temperature dependent\cite{Doi2013, DeGennes2002}, $\gamma=\gamma\left(T\right)$, a small local temperature-gradient suffices to generate a considerable propulsion. Using light, we heat up the light absorbing particle to establish a local temperature gradient, $\vec{\nabla} T$, at the vicinity of the fluid droplet. The temperature gradient generates a surface-tension gradient proportional to the thermo-capillary coefficient, $\beta \equiv \frac{\partial \gamma} {\partial T}$. The velocity, $\vec{v}$, of a fluid droplet in a temperature gradient is given by\cite{Young1959}:
\begin{equation}
\vec{v} = -\left[\frac{1}{2\eta_f + 3\eta_w}\right]\frac{\rm{D}}{2+\kappa_w/\kappa_f}\beta \vec{\nabla} T,
\label{eqThermoCapillarySwimmer}
\end{equation}
with $\rm{D}$ being the droplet diameter, $\eta_i$ is linear viscosity, and $\kappa_i$ the thermal conductivity (index $i$ indicates $w$-ater or $f$-luid droplet). Note that for high droplet viscosity, $\eta_f\gg\eta_w$, the swimming speed is dictated by the internal flow inside the fluid droplet, $v \propto \beta \nabla T /\eta_{f}$. We independently measured both the heating of the light absorbing particle and the thermo-capillary coefficient of the fluid droplet, finding numerical values consistent with existing literature \cite{Sloutskin2005, Baroud2007, Robert2008} (see Figs. \ref{figThermoCapillary}, \ref{figLaserHeating}, and section \ref{supSwimmerModel} in Supplementary Information for additional informatoin). By synthesising a family of swimmers made from a homologous series of alkanes, we observe a decrease in swimming speed, $v$, with increasing internal viscosity, $\eta_{f}$, as required by an {\it internal} drive (Fig.\ref{figSingleSwimmer}f). When accounting for the reduced mobility of particles near a solid surface \cite{Happel1983}, we find quantitative agreement between measured swimming speed, and the light driven thermo-capillary model for swimmers of different compositions (see Fig.\ref{figSingleSwimmer}f) with no fitting parameters.

\begin{singlespace}
\end{singlespace}

An individual swimmer can be oriented using an external magnetic field (through the small remanent moment of the superparamagnetic light absorbing particle) as shown in Fig.\ref{figSingleSwimmer}d,e. Fixing the orientation allows direct measurement of the swimming speed. Decoupling propulsion from random motion we find a persistent time of $\tau \approx 15\; \rm{s}$ for the free swimmer (Fig.\ref{figSingleSwimmer}a and section \ref{supPRW} in Supplementary Information). With increasing light intensity, individual swimmers transition from Brownian to ballistic (Fig.\ref{figSingleSwimmer}a) and can move up to 3 body lengths per second ($15 \;\mu\rm{m}/s$). Here we focus on the slower swimming speeds, $v\le5\;\mu\rm{m}/s$, corresponding to a P\'eclet Number, $P_e$, of up to $200$. In the presence of other swimmers, the individual swimmer remains active, but with dramatically different dynamics.

\begin{singlespace}
\begin{figure*}[h!t]
\includegraphics[width=1\textwidth]{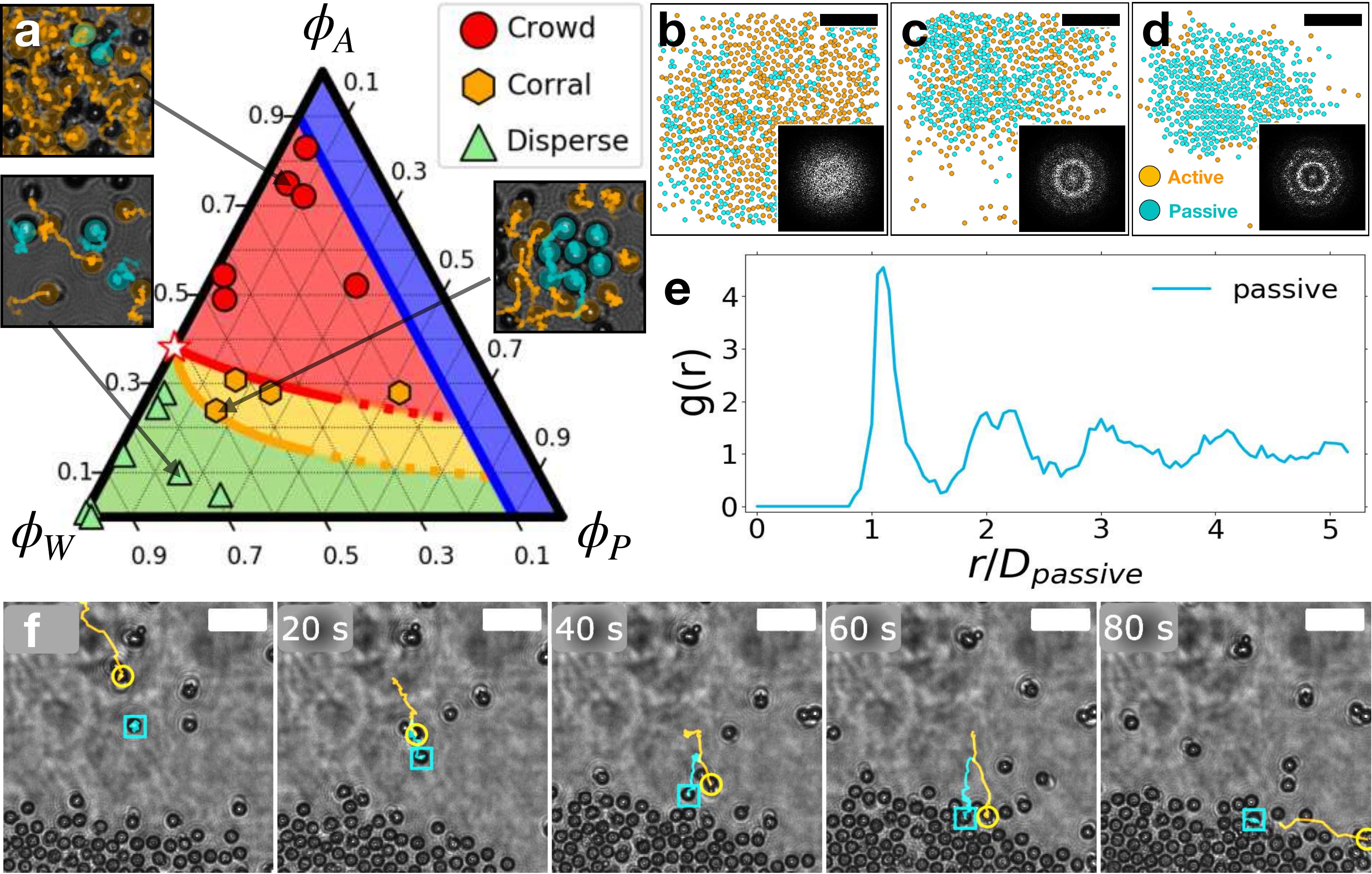} 
\caption{A swarm of active swimmers cooperatively corrals passive particles. {\bf a} a ternary phase diagram of active ($\phi_A$) passive ($\phi_P$) and the surrounding water bath ($\phi_W$) depicts the resulting dynamics of a mixture of passive particles and active swimmers, given their initial area fraction. Green triangles show that at low overall-particle area fraction (large $\phi_W$), the active swimmers quickly disperse. Red circles show a region where the active particles crowd once activated and form a persistent dense phase. Between the red and the green regions, $1-4\phi_P<\phi_A/\phi^*<1-\frac{\sqrt{12}}{\pi}\phi_P$, a yellow region emerges, where orange hexagons signify experiments where active particles spontaneously corral the passive particles ($\phi^*$ is denoted by a star). Red and orange curves are theoretical predictions for the onset of the corralling  state extrapolated for the small $\phi_P$ limit. Thumbnails show snapshots from corresponding points on the diagram (orange and cyan label active and passive particles respectively). Blue region indicates an overall particle area fraction greater than 0.91 (hexagonal packing). {\bf b}-{\bf d} snapshots taken at time 0 mins ({\bf b}) 5 mins ({\bf c}) and 30 minutes ({\bf d}) show the global dynamics of the active particles (orange) as they corall the passive particles (cyan) (see Supplementary Video 6). Insets are the structure factor, $S\left(\vec{q}\right)$,  of the passive particles, showing the evolution towards hexagonal order in their packed arrangement. {\bf e} the pair correlation function has peaks at 1, 2, 3, and 4 diameter, with $1-\sqrt3$ splitting around 2 indicative of ordered hexagonal packing. {\bf f} snapshots  at 20 seconds intervals show the corralling process which proceeds through entrainment events, where an active particle (orange) deposits a passive particle (cyan) at the dense region. After deposition, the active particle leaves (see Supplementary Video 2). Scale bars: {\bf b}-{\bf d} $30\;\mu\rm{m}$; {\bf f}: $10\;\mu\rm{m}$.}
\label{figCorralling}
\end{figure*}
\end{singlespace}

The individual swimmer motility is characterized by a significant slow down once entering a region of high local concentration (see Fig.\ref{figMIPS}a and Supplementary Video 1). It is known that the speed-concentration dependence of the individual swimmer, $v\left(\phi\right)$, can be used to predict the critical concentration, $\phi^*$, at which the whole swarm tends to form a persistent dense region\cite{Tailleur2008}. The trajectory of a single swimmer as it samples a heterogeneous density field is shown in Fig.\ref{figMIPS}a and Supplementary Video 1. The long lived crowded state is characterized by the co-existence of dense and dilute regions, with corresponding peaks in the density probability distribution (Fig.\ref{figMIPS}b). By measuring the local, instantaneous area fraction, $\phi$ (using Voronoi tesselation of surrounding swimmers, see section \ref{supVoronoi} in Supplementary Information), we find that a swimmer maintains its nominal speed at low concentrations, and experiences a sharp decrease in motility inside the dense region (Fig.\ref{figMIPS}b). The swimmer regains its nominal speed once emerging from the other side of the dense region. It is theoretically predicted that when the decrease in speed with concentration is sufficiently sharp, $\frac{\partial v}{\partial \phi}|_{\phi = {\phi^*}}\le-\frac{v}{\phi}$, concentration gradients appear at steady state in a process known as Motility Induced Phase Separation \cite{Tailleur2008}. The critical concentration we anticipate from $v\left(\phi\right)$ if Fig.\ref{figMIPS}b is $\phi^*\approx 0.38$, consistent with numerical predictions found in the literature \cite{Tailleur2008,Fily2012,Redner2013,Stenhammar2015,Cates2015b,Solon2015}. Indeed, the swarm's dynamics changes dramatically above this critical concentration. Below $\phi^*$ the initial population, $N_0$, quickly disperses, with the lion's share of the swimmers, $\Delta N$, leaving the observed field of view $\Delta N \approx N_0$. By contrast, at concentrations above $\phi^*$ the swarm maintains a long lived crowded huddle, with only a small fraction of the swimmers leaving, $\Delta N \ll N_0$ (see Fig.\ref{figMIPS}d). As in dense bacterial colonies \cite{Wensink2012,Secchi2016,Xu2019}, the crowded swimmer phase is liquid, characterized by flow correlations much larger than the individual, yet displays a turbulent structure (Fig.\ref{figMIPS}c), allowing internal re-arrangement, a key ingredient for their cooperativity.

As the swimmers remain active even at high concentrations, they cooperate to corral passive particles (Supplementary Video 6). Figures \ref{figCorralling}b-d show the time evolution of a heterogeneous mixture of active and passive particles, where the passive particles are compressed by active particles into a dense phase with hexagonal order, as seen in the pair correlation function, $g\left(r\right)$, and the structure factor, $S\left(\vec{q}\right)$ in Fig.\ref{figCorralling}b-e (see section \ref{supSofQ} in Suuplementary Information). Microscopically, the corralling builds-up through series of entrainment events, similar to those observed with microalga \cite{Jeanneret2016} where active particles chaperon passive particles to the dense region. The passive particles are then deposited while the active particles depart (Fig.\ref{figCorralling}f and Supplementary Video 2). A ternary phase diagram (Fig.\ref{figCorralling}a) summarizes the different dynamics observed given the initial area fractions of the active particles $\phi_A$, passive particles, $\phi_P$, and the surrounding water phase, $\phi_W$. When the initial concentration is less than dense packing (blue region $\phi_A+\phi_P\le0.907$), we identify three characteristic behaviours: active particles quickly disperse at low particle area fraction (green region); a long lived dense active state is formed at higher concentrations (red region); at intermediate particle concentrations, the active particles corral the passive particles (yellow region).

A simple geometrical construction captures the corralling phase boundaries in the phase diagram. Introducing passive particles reduces the total area available for the active particles, increasing their effective area fraction, $\phi^{\rm{eff}}_A$. The increased effective area fraction may surpass the critical concentration, $\phi^*$, and trigger the crowding dynamics. The extent of this effect however depends on the spatial arrangement of the passive particles. When randomly dispersed each passive particle occupies much more room as it carries a corona of excluded area. When the passive particles are packed in a lattice (as in Figs.\ref{figCorralling}d,e), their excluded areas overlap and they effectively occupy a smaller area. Therefore we expect the effective area fraction of the active particles to reduce when the passive particles are hexagonally packed (HP), $\phi_A^{P,\rm{rand}}>\phi_A^{P,\rm{HP}}$. For circular passive particles in 2D we can estimate the effective area fraction of the active particles for the two cases as, 
\begin{equation}
\phi^{\rm{eff}}_A \approx 
	\begin{cases} 
		\phi_A/\left(1-4\phi_P\right) & \rm{passive\; are\; random}  \\
		\phi_A/\left(1-\frac{\sqrt{12}}{\pi}\phi_P\right) &  \rm{passive\;  are\; HP}
	\end{cases}
	\label{eqCorral}
\end{equation}
(see Supplementary Information section \ref{supExcludedVolume} for full derivation). The red and orange curves in Fig.\ref{figCorralling}a respectively represent the point at which the effective area fraction of the active particles is equal to their critical area fraction for crowding, $\phi_A^{P,\rm{rand}}=\phi^*$ and $\phi_A^{P,\rm{HP}}=\phi^*$. The corralling region is found between those two curves.

The following picture arises: when active and passive particles are homogeneously mixed, the passive particles can trigger the crowding of the active particles at concentrations lower than the critical concentration of the pure active system, $\phi^*$. Once the active particles complete the corralling task, they disperse, as the area gained by the overlapping excluded areas makes them sub-critical. A similar equilibrium process is the depletion interaction \cite{Doi2013,Kardar2007}, where the diffusion of many small particles (depletants) can drive fewer, larger particles to crystallize. There is one important distinction: even in the absence of passive particles, active particles have an intrinsic critical clustering concentration because of their activity. Our analysis shows qualitatively similar results to previous theoretical work on mixed active/passive populations \cite{Weber2016,Grosberg2015,Stenhammar2015,Redner2013}, but may prove more general as it relies on a generic geometrical argument. 

In this work we have introduced a synthetic route and a driving strategy for a new family of micro-swimmers that are propelled solely by light --- a colloidal chimera of a light absorbing particle bound to a fluid droplet. An individual swimmer's propulsion mechanism is quantitatively captured by a thermo-capillary drive. The swimmers remain active at high area fractions, $\phi_A>0.5$, and can be used as an experimentally tuneable model system to study a host of phenomena in dense active matter such as 2D turbulence in a compressible fluid, previously seen in bacterial colonies\cite{Xu2019, Dunkel2013, Wensink2012, Zhang2010c, Dombrowski2004}. We found that at low concentrations swimmers tend to quickly disperse, but above a critical concentration form a long lived dense phase. We identify a depletion-like interaction where the excluded volume of passive particles in random or close packed configurations can change $\phi^*$ of the active particles resulting in passive particle crystallization. As the mechanism is geometric, it may be applicable generally from living organisms to robotic swarms. As our optically powered Marangoni swimmers are largely agnostic to their precise chemical environment, we were able to graft DNA on their surface. Micro-swimmers augmented by DNA nanotechnology \cite{Seeman2015,BenZion2017} can prove useful in the self-assembly of active tissue, mimicking biofilms, in the making of future, {\it functional} active matter.

%

\bibliographystyle{naturemag}

\subsection*{Acknowledgements}
We thank N. Oppenheimer, O. Dauchot,  A. M. Leshansky, T. Goldfriend and C. P. Kelleher for insights. This research was primarily supported by the Center for Bio-Inspired Energy Sciences, an Energy Frontier Research Center funded by the DOE, Office of Sciences, Basic Energy Sciences, under award no. DE-SC0000989, and ), and the Diversity Undergraduate Research Initiative (DURI) New York University.

\section*{Methods}\label{supMethods}
\section*{Swimmer Synthesis and Sample Preparation}\label{supSamplePreparation}
Hybrid dimers were synthesized by binding a $3\;\mu\rm{m}$ light absorbing particle (M280 BANGS LAB) to a fluid oil particle (see list in section \ref{supSwimmerModel}) using the previously introduced Mix-And-Match strategy\cite{BenZion2020}. Briefly, fluid droplets were made through membrane emulsification where oil with $0.1\%$ v/v SPAN80 (SIGMA) is pressurized through a porous membrane forming a stable oil in water emulsion (stable for months at room temperature). Typical particle size was $3\;\mu\rm{m}$. M280 particle and fluid droplets are stoichiometrically mixed, destabilized, quenched, and purified using density gradient step centrifugation. Dimers were found to be stable for months when stored in a refrigerator at $4^o\rm{C}$. Samples were prepared for imaging by loading a suspension to a pre-treated glass capillary (channel height $0.10$ mm, width $2.00$ mm Vitrotubes W5010-050), that were plasma cleaned (SPI supplies Plasma Prep II), and passivated through vapour deposition of hexamethyldisilazane (SIGMA). Loaded capillaries were placed on a clean microscope glass slide, and sealed on their ends with UV curable resine (LOON OUTDOORS UV CLEAR FLY FINISH).

\section*{Experimental Setup}\label{supExperimentalSetup}
Imaging was done using bright field on a home built microscope coupled to a laser source. A commercial light emitting diode ($\lambda = 505$ nm THORLABS) with a diffuser (ground glass N-BK7 600 grit, THORLABS), a condenser and an iris, to achieve Köhler illumination. Scattered light was picked up by the microscope objective (HCX PL APO 40x NA = 0.85, LEICA), and a tube lens (B\&H), and detected by a digital camera (DCC1545M, IMAGING SOURCE) and acquired using commercial video recording software (IC CAPTURE, IMAGING SOURCE). A heating beam was introduced on a separate optical path (see Fig.\ref{figExperimentalSetup} in Supplementary Information). A $1064$ nm laser beam (YLR-10-1064-LP, IPG Photonics) was passed through a zero order half plate (WPH05M-1064 Thorlabs) and contracted using a custom Galilean telescope to achieve a $\sim 300\;\mu$m beam (see Supplementary Information). The laser beam was introduced into the sample using a polarizing beam splitter (PBS CM1-PBS253 THORLABS) and its intensity at the sample was controlled by a combination of the electronic laser head controller and by adjusting the half-plate, and was measured using an optical power meter (PM100D power meter, with S175C sensor, THORLABS). In order to eliminate laser intensity before the camera, stained glasses (FGS900S, THORLABS) were stacked after the objective. 

\subsection*{Supplementary Videos}
\begin{description}[itemsep=2.45mm]
\item \href{https://youtu.be/ZEfo46HDgSU}{S1 - Swimmer's activity inside the dense phase.} 
\item \href{https://youtu.be/Lrz_xtDfMm4}{S2 - Entrainment mediated corralling.} 
\item \href{https://youtu.be/RI0CI5ESrjU}{S3 - Quick dispersing at low density.}
\item \href{https://youtu.be/7XR24PIX0hI}{S4 - Schooling and anti-schooler at intermediate density.}
\item \href{https://youtu.be/j6dBq0F12nY}{S5 - Internal flow inside the dense phase.}
\item \href{https://youtu.be/ZYLvVHBA0KY}{S6 - Corralling of passive particles.}
\end{description}

\section{Light Driven Thermo-Capillary Swimmer Model}\label{supSwimmerModel}
\subsection{Fluid Droplet in a Temperature Gradient}
For the thermo-capillary swimmers presented, the temperature gradient is given by the temperature profile around the hotter light absorbing particle when it is exposed to radiation $ \vec{\nabla} T = - \frac{P_0}{4\pi \kappa_w \rm{R}^2} {\bf \hat{e}_r}$, where $P_0$ is the heating power at the particle, and $R$ the distance from its center, along the radial direction ${\bf \hat{e}_r}$. The power absorbed by the particle is given by $P_0=\sigma J$, where $J$ is the flux and $\sigma$ the particle's absorption cross section (see section \ref{supLaserHeating} for measurement of $\sigma$). As on the length scale of the swimmers, heat transport is dominated by high thermal diffusivity, the temperature profile is assumed to move with the swimmer. Since the droplet is intimately bound to the particle, we approximate the temperature gradient to be constant with its value at the center of the droplet, allowing us to use Eq.\ref{eqThermoCapillarySwimmer}. To quantitatively verify that indeed this is the correct swimming mechanism, we directly measured the thermo-capillary coefficient, $\beta$ (see section \ref{supThermoCapillary}), and use literature data for the thermal conductivies and viscosities of the different swimmers species (see Table \ref{tblOils}). The swimmer's proximity to the wall, $\delta \approx 0.2$ nm (estimated from its gravitational height), leads to enhanced drag, estimated to increase by a factor of $\frac{8}{15} \rm{log}\left( \frac{\rm{D}}{2\delta}\right)$ for a sphere moving along a solid wall in Stokes flow \cite{Happel1983}. 

\begin{center}
\begin{table}[H]
\centering
\begin{tabular}{ l c c }
 Material & $\kappa_f \left(\frac{\rm{W}}{\rm{K\;m}}\right)$ & $\eta_{\rm{f}} \left( \rm{m Pa s} \right)$ \\ 
 n-decane & 0.15  & 0.93 \\ 
 n-dodecane & 0.14  & 1.5 \\ 
 n-tetradecane & 0.14  & 2.3 \\
 n-hexadecane & 0.14  & 3.5 \\
 mineral oil & 0.17 *& 13 * \\
\end{tabular}
\caption{\label{tblOils} Material properties of different fluid droplets were taken from the literature \cite{Mustafaev1973,Calado1983,Tanaka1988,Schmidt2014}.  According to manufacturer. }
\vspace{-1cm}
\end{table}
\end{center}

\subsection{Thermo-Capillary Measurement}\label{supThermoCapillary}
\begin{figure}[h]
\centering
\includegraphics[width=0.46\textwidth]{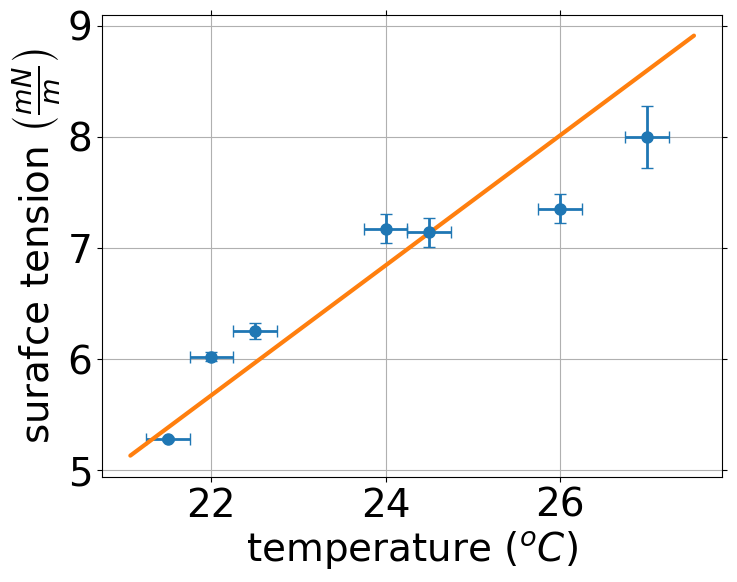} 
\caption{Linear thermo-capillary coefficient measurement.}
\label{figThermoCapillary}
\end{figure}

Surface tension measurements were carried out using a commercial optical tensiometer (THETA ONE ATTENSION) by hanging a deionized water drop ($18\;\rm{M}\Omega\rm{cm}$, TOC = $8$ ppb, MILLIPORE MILLI-Q) in a n-dodecane (ALPHA AESAR) bath with $0.1\%$v/v SPAN-80 (SIGMA) inside a glass cuvette (THERMOFISHER) and fitting the profile of the pendant drop to the Young-Laplace model \cite{DeGennes2002}. The cuvette was wrapped with resistive heating wire to ensure a homogeneous temperature, monitored using a thermometer (BS4 59-7569 Monitoring Thermometer) with a thermo-couple sensor (IT-21 microprobe) placed in close proximity to the droplet. Surface tension readings were taken after pendant drop was allowed to equilibrate at each temperature. The surface tension was then measured for multiple droplets, and their average reading was taken. Temperature was varied between $21$-$27^o$C and with a surface tension increase from $5.2\;\frac{\rm{mN}}{\rm{m}}$ at room temperature, to $8\;\frac{\rm{mN}}{\rm{m}}$. The linear thermo-capillary coefficient is found from linear regression (see Fig.\ref{figThermoCapillary}) to be $\beta = 0.58\pm0.02\;\frac{\rm{mN}}{\rm{^oCm}}$, quantitatively consistent with known values in the literature \cite{Sloutskin2005, Baroud2007, Robert2008}. 

\subsection{Laser Heating Power Balance}\label{supLaserHeating}
\begin{figure}[h]
\centering
\includegraphics[width=0.5\textwidth]{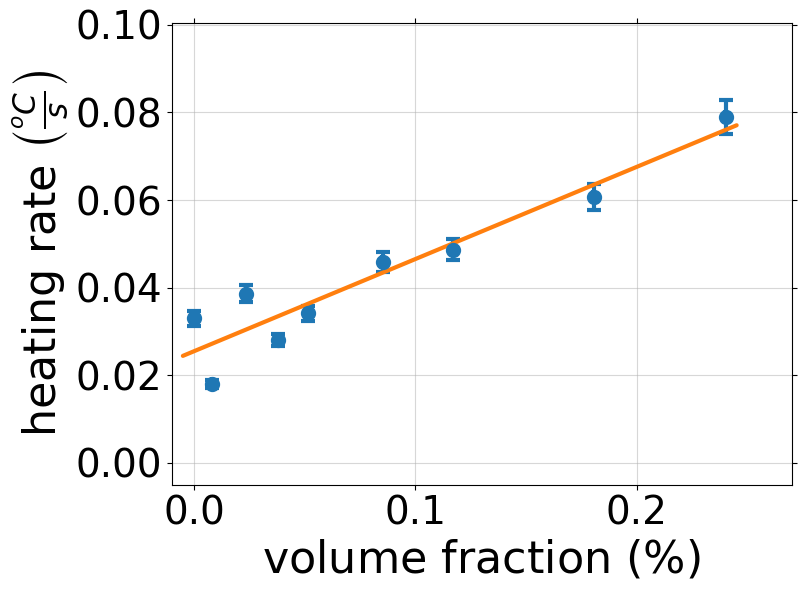} 
\caption{Thermal heating due to light absorption}
\label{figLaserHeating}
\end{figure}

Heating measurements were carried out by tracking the temperature increase rate, $\frac{\partial T}{\partial t}$, (BS4 59-7569 Monitoring Thermometer) with a thermo-couple sensor (IT-21 microprobe) placed inside a standard glass cuvette (THERMOFISHER length $L=1$ cm) filled with a suspension of $D=3\;\mu$m light absorbing particles (M280 Bangs Lab) in deionized water ($18\rm{M}\Omega\rm{cm}$, TOC=$8$ ppb, MILLIPORE MILLI-Q), when irradiated with a collimated $1064$ nm laser beam (YLR-10-1064-LP, IPG Photonics) while stirring using a magnetic stir bar to ensure rapid thermal homogenization. The total light intensity of the beam, $P=1.3$ W was measured using an optical power meter (PM100D power meter, with S175C sensor, THORLABS), and  temperature increase rate was found from the temperature vs. time slope at early times. In order to extract the particle's absorption cross section, $\sigma = \pi \left(\frac{D}{2}\right)^2\sigma_0$ (where $\sigma_0$ is the differential cross section for absorption), the measurements were taken at increasing volume fraction, $\Phi$ ($0-0.4\%$v/v), eliminating the large but constant contribution to the heating, $W$, by the water bulk. Assuming dilute  suspension ($\Phi\ll1$), small temperature gradients from ambient ($t\ll 1$ min), and fast homogenization (stirring), the light absorbing particles contribution to the heating is given by the slope of 
\begin{equation}
\frac{\partial T}{\partial t} = W+\frac{\frac{3}{2}\frac{L}{D}P}{C_P}\sigma_0\Phi,
\label{heatingPerParticleDimensionless}
\end{equation}
where the heat capacity of the fluid, $C_P=c_p m_{\rm{H_2O}}$ is dominated by the water, and the differential cross-section is found to be $\sigma_0 = 5\%$, (see Fig.\ref{figLaserHeating}).

\subsection{Beam Size Calibration}\label{supBeamSize}
\begin{figure}[h]
\centering
\includegraphics[width=0.4\textwidth]{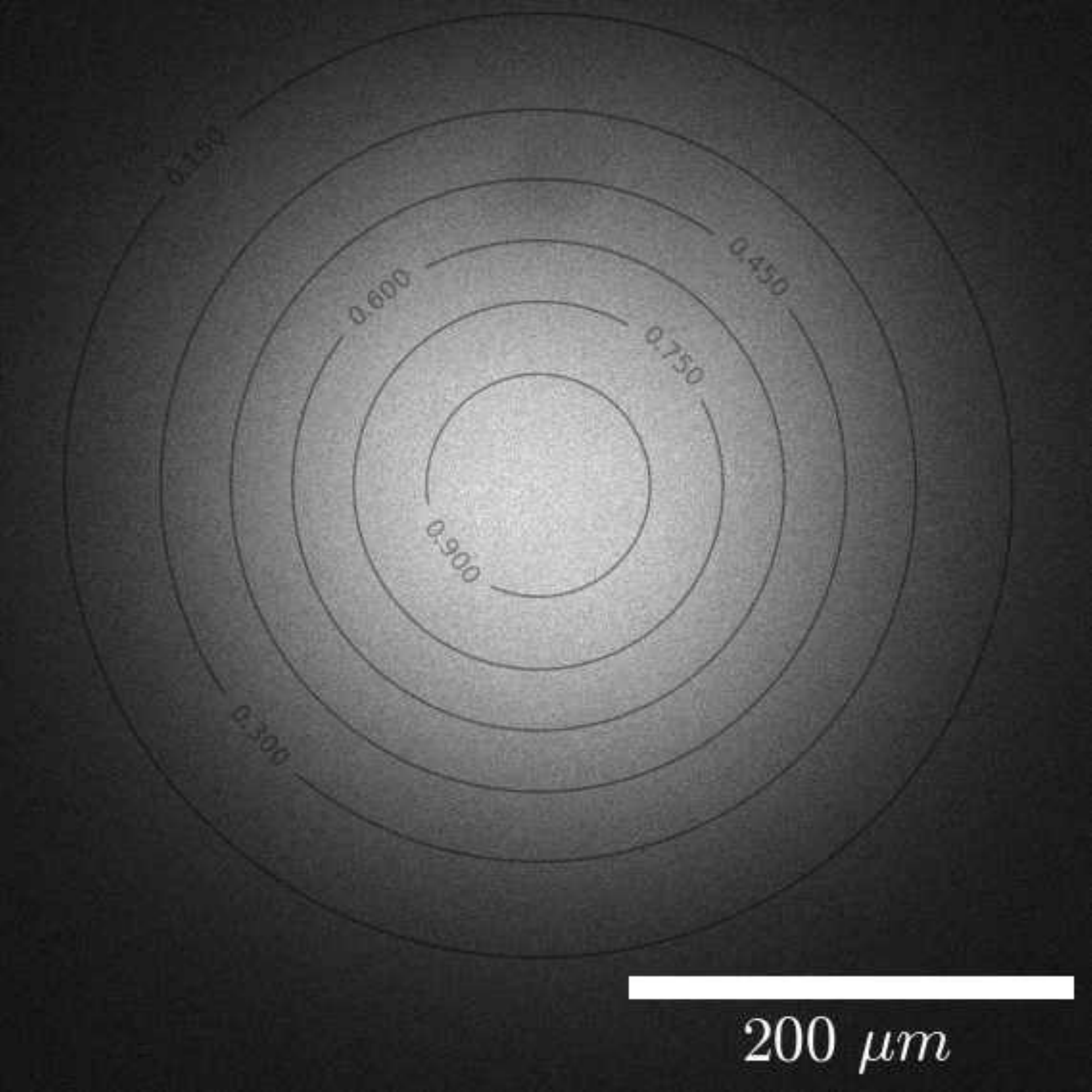} 
\caption{Broad field image of laser beam (contours show relative power given by Gaussian fit). Experiments were performed within a field of view of $\sim100\;\mu\rm{m}$ at the center of the beam where power spatial variation is smaller than 10\%.\vspace{-0.5cm}}
\label{figBeamSizeCalib}
\end{figure}

The beam profile was measured by exposure to infrared visualizing shavings (980nm-1064nm IR Infrared Laser Visualizer) fixed inside a UV curable resin (LOON OUTDOORS UV CLEAR FLY FINISH) between a microscope slide and a cover-slip. Laser beam intensity, $\rm{I}$, was measured at the objective using an optical power-meter (PM100D power meter, with S175C sensor, THORLABS). The visible light emitted from the IR visualizer was imaged using a lower magnification objective (Leica HC Pl Fluotar 10X), to create the profile seen in Fig.\ref{figBeamSizeCalib}). The spatial distribution of the beam $J\left(R\right)$ was then fitted to a two dimensional Gaussian $ J=J_0e^{-\rm{R}^2/\sigma^2} $, with $R$ the radial distance from the Gaussian center, $J_0 = \frac{\rm{I}}{\pi \sigma^2}$ being the flux at the center of the beam, and $2\sigma = 310\;\mu\rm{m}$ is the fitted beam diameter (see Fig.\ref{figBeamSizeCalib}).

\begin{figure}[h]
\centering
\includegraphics[width=0.5\textwidth]{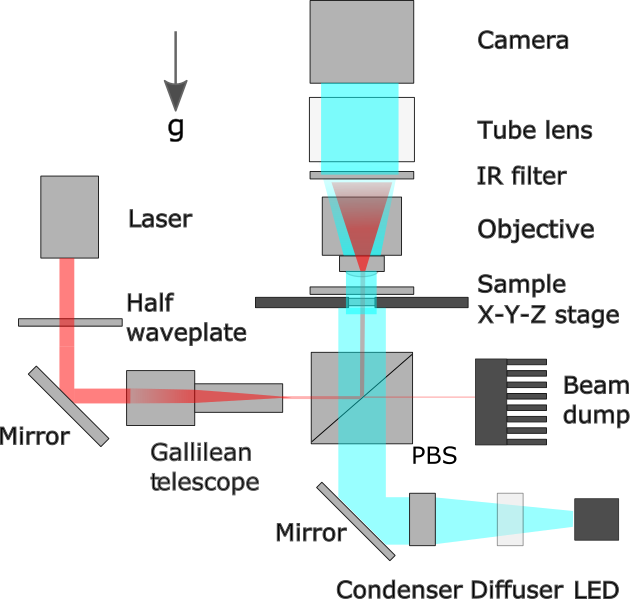} 
\caption{Schematics of experimental setup.}
\label{figExperimentalSetup}
\end{figure}

\section{Persistent Swimmer Model}\label{supPRW}
Individual swimmer's dynamics were modelled as a persistent random walk with speed $v$, persistent time $\tau$, and a diffusion constant $D_0$. The mean square displacement, $\left<\Delta L^2\right>$, of such a swimmer can be found in \cite{Howse2007}:

\begin{equation}
\left<\Delta L^2\right> = 4D_0\Delta t+\frac{v^2\tau^2}{2}\left[\frac{2\Delta t}{\tau} +e ^{-2\Delta t/\tau}-1 \right].
\label{eqPRW}
\end{equation}
Which for the short time limit ($\Delta t \ll \tau$) is ballistic, $\left<\Delta L^2\right> \approx v^2\Delta t^2 \propto \Delta t^2 $, and for longer lag times ($\Delta t \gg \tau$) behaves diffusively with an enhanced diffusion constant, $\left<\Delta L^2\right> \approx \left(4D_0 +v^2\tau \right)\Delta t \propto \Delta t^1$. Swimmer videos were preprocessed using ImageJ \cite{Rueden2017}, and the subsequent particle motion tracked and analyzed using Trackpy package on Python \cite{Allan2019}. Fig.\ref{figSingleSwimmer}{\bf a} in the main text shows the Brownian ($\propto \Delta t^1$) to ballistic ($\propto \Delta t^2$) transition for increased flux. Swimming speed was independently measured by imposing their orientation using a custom built vector magnet made of 3 Helmholtz pairs, generating a constant homogenous magnetic field ($\sim1$mT) in the imaging plane. In the presence of a magnetic field, the swimmer maintains a fixed direction (see Fig.\ref{figSingleSwimmer}{\bf d}), allowing extraction of the swimmers speed, $v$, by measuring displacement over time, which is found to be linear with flux (as shown in Fig.\ref{figSingleSwimmer}{\bf b}) and is expected by the model (Eq.\ref{eqThermoCapillarySwimmer}). Note that the swimming direction is not necessarily along the magnetic field as the dimmer orientation is decoupled from the magnetic dipole (for example see Fig.\ref{figSingleSwimmer}{\bf d}). We find that the reorientation time of the swimmer is then $\tau \approx 15$ s, corresponding to a persistence length of up to $70\;\mu$m with a Peclet number $Pe\approx 200$ using the definition found in the literature \cite{Marchetti2013}. As our microscope imaging is done from the top, with the laser introduced from the bottom (see Fig.\ref{figExperimentalSetup}), swimmers are subjected to radiation pressure and swim on the ceiling of the capillary, explaining the small deviation from linearity at low fluxes in Fig.\ref{figSingleSwimmer}b. We did however perform similar experiment on an inverted version of our apparatus (that is inverted imaging and laser introduced from above), finding similar swimmers dynamics.

\section{Local Density Measurement}\label{supVoronoi}
Local instantaneous density measurements were done by locating the two dimensional coordinates of all the particles in each frame $\left(x_i,y_i,t_i\right)$ \cite{Allan2019}, around which a Voronoi tessellation was constructed using standard Python routines. The local area fraction of each particle $\phi_i$ is found by dividing the area of each particle, $a_i$ by the area, $A_i$ of its Voronoi cell $\phi_i \equiv \frac{a_i}{A_i}$. Histograming these values over 2600 frames gives the two population densities found in Fig.\ref{figMIPS}b. The area fraction field, $\phi\left(x,y,t\right)$ is then found piecewise from the local area fraction of each particle $\phi\left(x,y,t\right) = \sum_i \delta\left(x-x_i\right)\delta\left(y-y_i\right)\delta\left(t-t_i\right)\phi_i$ (where $\delta$ is the Dirac delta function). The single swimmer speed as a function of concentration $v\left(\phi\right)$ is obtained by tracking a swimmer's displacement, $\left| \Delta r \right|$, moving in the dynamic environment. Average displacement as a function of concentration $\left<\left|\Delta r\right|\right>\left(\phi\right)$ is then found by averaging displacements within density bins of $6.7\%$ area fraction within time intervals of $5$ seconds from which points in Fig.\ref{figMIPS}b are measured. Error bars are the standard deviation of the displacement. A prediction for critical crowding area fraction, $\phi^*$, is the found by fitting the measured displacement-concentration curve $\Delta \left| r\right| \left(\phi\right)$ to a sigmoid $f\left(\phi\right) =\frac{B}{1+e^{\left(\phi-\phi^*\right)/\alpha}}+C$ ($\sigma$, $B$, and $C$ are constants) and the critical crowding concentration can be found from when satisfying the theoretically expected relation \cite{Tailleur2008} of the derivative: $\frac{\partial f}{\partial \phi}= -\frac{f}{\phi}$.

\section{Critical Crowding Concentrations}\label{supManySwimmers}
In order to find the critical area fraction, $\phi^*$, above which swimmers form a crowded phase, a series of experiments were carried out initiating swimming in regions with a homogeneous swimmer distribution with an initial mean area fraction, $\phi_0$ calculated by the total area of the observed particles, $\sum^{N_0}_{i=1} a_i$, over the area of the field of view ($A_{\rm{FOV}}$ kept constant throughout the experiment) $\phi_0 \equiv \frac{1}{A_{\rm{FOV}}}\sum^{N_0}_{i=1} a_i$, where $N_0$ the initial number of particles. As the collective dynamics evolve, with the number of swimmers inside the FOV is monitored as a function of time $N\left(t\right)$, given the evolution of the relative number $\frac{N\left(t\right)}{N_0}$ is found in Fig.\ref{figMIPS}d.

\section{Velocity Correlation Function from Particle Image Velocimetry}\label{supPIV}
The normalized velocity correlation function, 
\begin{equation}
\left<\vec{V}\left(0\right)\cdot \vec{V}\left(\vec{R}\right)\right>_N =\frac{1}{\left<\vec{V}\left(\vec{r}\right)^2\right>_r} \left<\vec{V}\left(\vec{r}\right)\cdot \vec{V}\left(\vec{r}+\vec{R}\right)\right>_r,
\label{eqVCF}
\end{equation} was found from the auto correlation of the velocity field, $\vec{V}\left(\vec{R}\right)$ ($\left<\right>_r$ denotes spatial average). The velocity field $\vec{V}\left(\vec{R}\right)$ was found using {\it OpenPIV} python library \cite{Liberzon2020} from pairs of frames $1$ second apart taken from within the dense phase for the duration of the experiment. The VCF is computed component-wise in Fourier space $\tilde{V_i}\left(\vec{k_i}\right)$ (index $i$ is for each coordinate $i\in\left[x,y\right]$) then using convolution theorem, converted to real space from the inverse Fourier transform of $\tilde{VCF}\left(\vec{k}\right) = \left|\tilde{V_x}\right|^2\left(\vec{k}\right)+ \left|\tilde{V_y}\right|^2\left(\vec{k}\right)$, with angular averaging to extract the radial dependence.

\section{Spatial Correlations of Passive Particles}\label{supSofQ}
Structure factor for the passive particles, $S\left(\vec{q}\right)$, was defined by
\begin{equation}
S\left(\vec{q}\right) = \frac{1}{N} \left|\tilde{\rho}\left(\vec{q}\right)\right|^2,
\label{eqSofQ}
\end{equation}
where $\vec{q}$ is the wave vector, and $\tilde{\rho}\left(\vec{q}\right)$ is the Fourier transform of the particle density, $\rho\left(\vec{r}\right)$, which we compute by plotting $\rho\left(\vec{r}\right)=\sum_i^NF\left(\vec{R_i}\right)$, with $\{ \vec{R_i}\}$ being the set of experimentally located particle coordinates, and $F$ being a Gaussian form factor that reduces short wave length noise. 
The radial pair correlation function, $g\left(r\right)$ in the same figure was calculated by directly histogramming the distances, $r$, measured any two pairs of passive particles, with constant binning of $200$ nm ($0.067$ particle diameter).

\section{Corralling Through Excluded Area}\label{supExcludedVolume}
The collective dynamics were measured by choosing a field of view with a rather homogenous particle distribution, turning the light on, and tracking the dynamics. We define the three states as follows: dispersing state where after turning on the light source, most of the particles leave the field of view (FOV); the crowding state where after turning on the light still most of the swimmers are still inside the FOV; and the corralling state where the major entity left in the FOV is the passive particles. Points on Ternary phase space are defined as described in section \ref{supManySwimmers}, by measuring the initial area fraction of passive $\phi_P$ and active $\phi_A$, and plotted using {\it python-ternary} package \cite{Harper2019}.

\begin{figure}[h]
\centering

\includegraphics[width=0.5\textwidth]{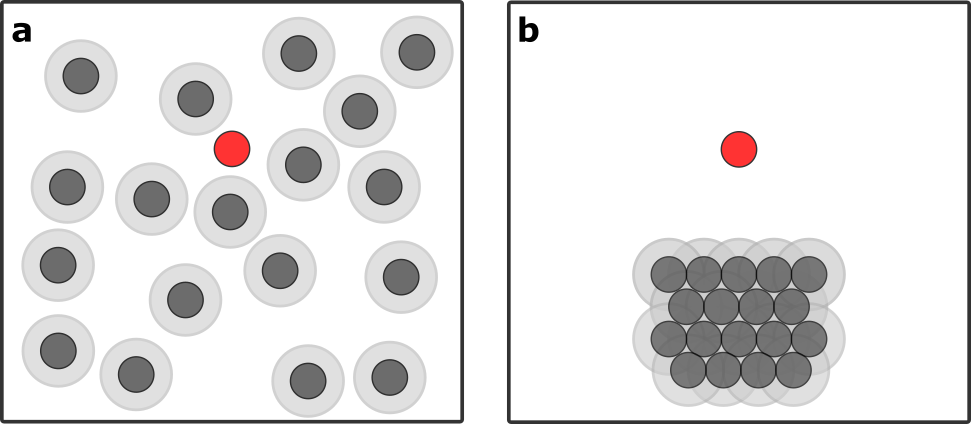} 
\caption{Geometry of effective area occupied by 18 passive particles when {\bf a} randomly distributed, or {\bf b} hexagonally packed. Light shaded areas are the regions excluded for the test particle (red). }
\label{figExcludedAreas}
\end{figure}

To derive the criterion for corralling we note that the area available for the active particles, $A$, is the total area, $A_T$, after accounting for the effective area taken by the passive particles $A^{eff}_P$

\begin{equation}
A = A_T - A^{eff}_P.
\label{eqCorra1l}
\end{equation}
Generally the area occupied by the passive particles depends on their spatial arrangement. Given that there are $N_A$ active ($N_P$ passive) particles, each of which occupies an area $a_A$ ($a_P$), their area fraction is defined as $\phi_A\equiv\frac{ a_A N_A}{A_T}$ ($\phi_P \equiv\frac{ a_P N_P}{A_T}$). The effective area fraction of the active particles is then defined as $\phi^{\rm{eff}}_A \equiv \frac{N_A}{A}$  giving

\begin{equation}	
\phi^{\rm{eff}}_A = \phi_A \Big/\left(1-\frac{A^{eff}_P}{A_T}\right),
\label{eqCorral2}
\end{equation}
which in general also depends on the spatial arrangement of the passive particles. The question thus becomes the following: for a given spatial arrangement of the passive particles, is the effective area fraction of the active particles super critical: $\phi^{\rm{eff}}_A>\phi^*$? We shall treat two cases of different spatial arrangement of the passive particles, Random, and  Hexgonally Packed (HP). When the passive particles are randomly distributed, additionally to their own area, $a_P$, each particle occupies a corona of excluded region, inaccessible to other particles (see Figure \ref{figExcludedAreas}a). For the simple case of equisized particles, this excluded area is 3 times that of the particle itself, $a^e_P = 3a_P$, making the effective area occupied by the passive particles 
\begin{equation}
A^{\rm{rand}}_P=4a_P N_P.
\label{eqRand}
\end{equation}
If however the passive particles are arranged in a HP crystal, their effective area is considerably smaller, $A^{\rm{HP}}_P = \frac{\sqrt{12}}{\pi} a_P N_P + O\left(\sqrt{N_P}\right)$ (as can be seen in Fig.\ref{figExcludedAreas}b). Here the excluded area becomes sub-extensive as it grows like the perimeter of the crystal, and for large enough crystals, we can approximate 
\begin{equation}
A^{\rm{HP}}_P \approx \frac{\sqrt{12}}{\pi} a_P N_P.
\label{eqHP}
\end{equation}
Plugging the two cases (Eqs. \ref{eqRand} and \ref{eqHP}) into Eq. \ref{eqCorral2} we recover the result in the main text (Eq.\ref{eqCorral}). This shows that there are combinations of concentrations of the active and passive particles, where the effective area fraction of the active particles, $\phi_A^{eff}$, can either be above or below the critical crowding concentration, $\phi^*$, depending on the arrangement of the passive particles. The system is expected to be in the crowding state when even if the passive particles were hexagonally packed, the effective area fraction of the active particles is still {\it above} critical, $\phi^{eff}=\phi_A \Big/ \left(1-\frac{\sqrt{12}}{\pi}\phi_P\right)>\phi^*$ (red region in phase diagram in Fig.\ref{figCorralling}a in the main text). Correspondingly, if the passive particles were randomly distributed, yet the effective area fraction of the active particles was still {\it below} critical, $\phi^{eff}_A\phi_A \Big/\left(1- 4\phi_P\right)<\phi^*$, the system is said to be in the dispersing regime (green shade in phase diagram in Fig.\ref{figCorralling}{\bf a}). Between these two regimes, the corralling state is defined (yellow region in Fig.\ref{figCorralling}{\bf a}).

\end{document}